\documentstyle[aps,preprint,tighten]{revtex}

\newcommand{\lsim}{\mathrel{\mathop{\kern 0pt \rlap
  {\raise.2ex\hbox{$<$}}}
  \lower.9ex\hbox{\kern-.190em $\sim$}}}
\newcommand{\gsim}{\mathrel{\mathop{\kern 0pt \rlap
  {\raise.2ex\hbox{$>$}}}
  \lower.9ex\hbox{\kern-.190em $\sim$}}}
\newcommand{\beq}    {\begin{equation}}
\newcommand{\eeq}    {\end{equation}}
\newcommand{\beqarr} {\begin{eqnarray}}
\newcommand{\eeqarr} {\end{eqnarray}}
\newcommand{\barr}   {\begin{array}}
\newcommand{\earr}   {\end{array}}
\newcommand{\no}     {\nonumber}

\newcommand{\mb}[1]  {\mbox{#1}}

\begin{document}

\preprint{
\begin{tabular}{r}
DFTT 48/98
\end{tabular}
}

\title{Compatibility of the new DAMA/NaI data on an annual modulation effect 
in WIMP direct search  with a relic neutralino in supergravity schemes}

\author{
\bf A. Bottino$^{\mbox{a}}$
\footnote{E--mail: bottino@to.infn.it,donato@to.infn.it,
fornengo@to.infn.it,scopel@posta.unizar.es},
F. Donato$^{\mbox{a}}$, N. Fornengo$^{\mbox{a}}$, 
S. Scopel$^{\mbox{b}}$\footnote{INFN Post--doctoral Fellow}
\vspace{6mm}
}

\address{
\begin{tabular}{c}
$^{\mbox{a}}$
Dipartimento di Fisica Teorica, Universit\`a di Torino and \\
INFN, Sezione di Torino, Via P. Giuria 1, 10125 Torino, Italy
\\
$^{\mbox{b}}$ Instituto de F\'\i sica Nuclear y Altas Energ\'\i as, \\
Facultad de Ciencias, Universidad de Zaragoza, \\
Plaza de San Francisco s/n, 50009 Zaragoza, Spain
\end{tabular}
}
\maketitle

\begin{abstract}
Recent results of the DAMA/NaI experiment for WIMP direct detection 
point to a possible annual modulation effect in the detection rate. 
We show that these results, when interpreted in terms of a relic
neutralino, are compatible with supergravity models. 
Together with the universal SUGRA scheme, we also consider SUGRA models where 
the unification condition in the Higgs mass parameters at GUT scale is relaxed. 

\end{abstract}  

\vspace{1cm}


\section{Introduction}

The new DAMA/NaI data (running period \# 2)\cite{dama2} 
provide a further  indication of 
a possible annual modulation 
effect in the rate for WIMP direct detection, already singled out 
by the same Collaboration using former data (running period \# 1) \cite{dama1}. 
In Ref. 
\cite{noi} we interpret the total sample of new and former data in terms 
of a relic neutralino in the framework of the Minimal 
Supersymmetric extension of the Standard Model (MSSM) \cite{susy}, 
by extending the analysis that we performed previously \cite {noi0} about the 
DAMA/NaI results of the running period \# 1. 
 The MSSM scheme represents a very versatile approach for 
discussing supersymmetric phenomenology at the Electro--Weak 
(EW) scale, and does not bear on too strong theoretical assumptions
at higher energies. In \cite{noi},  using MSSM, we prove 
 that the annual modulation data are quite compatible with 
a relic neutralino which may make up the major part of dark matter 
in the Universe and that some of the most relevant supersymmetric properties 
are explorable at accelerators in the near future.

In the present paper we show that the supersymmetric features,
 implied by the DAMA/NaI 
modulation data, are also compatible with more 
ambitious supersymmetry 
schemes, where the previous phenomenological model is implemented in 
a supergravity (SUGRA) framework, especially if the unification 
conditions, which are frequently imposed at the Grand  
Unification (GUT) scale, are appropriately relaxed \cite{arn}.

\section{SUGRA models} 

We remind here that the essential elements of SUGRA models 
\cite{sugra,ceop} are: 
a Yang--Mills Lagrangian, the
superpotential, which contains  all the Yukawa interactions
between the standard and supersymmetric fields, and  the soft--breaking
Lagrangian, which models the breaking of  supersymmetry.
Here we only recall the soft supersymmetry breaking terms
\begin{eqnarray}
&-{\cal L}_{soft}& =
\displaystyle \sum_i m_i^2 |\phi_i|^2
\no \\
&+&  \left\{\left[
A^{l}_{ab} h_{ab}^{l} \tilde L_a H_1 \tilde R_b +
A^{d}_{ab} h_{ab}^{d} \tilde Q_a H_1 \tilde D_b +
A^{u}_{ab} h_{ab}^{u} \tilde Q_a H_2 \tilde U_b +\mb{h.c.} \right] -
B \mu H_1 H_2 + \mb{h.c.}      \right\}
\no \\
&+& \displaystyle \sum_i M_i
(\lambda_i \lambda_i + \bar\lambda_i \bar\lambda_i)\;,
\label{eq:soft}
\end{eqnarray}
\noindent
where the $\phi_i$ are  the scalar fields, the $\lambda_i$ are the
gaugino fields, $H_1$ and $H_2$ are the two Higgs fields,
$\tilde Q$ and $\tilde L$
are the doublet squark and slepton fields, respectively,
and $\tilde U$, $\tilde D$ and
$\tilde R$ denote the $SU(2)$--singlet fields for the up--squarks,
down--squarks and sleptons. In Eq.(\ref{eq:soft}), $m_i$ and $M_i$ are the mass
parameters of the scalar and gaugino fields, respectively, and $A$ and
$B$ denote trilinear and bilinear supersymmetry breaking parameters,
respectively. The
Yukawa interactions are described by the parameters $h$, which
are related to the masses of the standard fermions by the usual
expressions, e.g. $m_t = h^t v_2$ and $m_b = h^b v_1$, where
$v_i = <H_i>$ are the v.e.v.'s of the two Higgs fields.

It is worth recalling that one
attractive feature of the model is the connection between
soft supersymmetry breaking and 
Electro--Weak Symmetry Breaking (EWSB), which would then be induced radiatively.

 It is customary to implement the supergravity  framework  
with some restrictive assumptions about unification at grand unification scale 
$M_{GUT}$:

     ~i) Unification  of the gaugino masses: 
        $M_i(M_{GUT}) \equiv m_{1/2}$,

     ii) Universality of the scalar masses with a common mass denoted by
     $m_0$: \hfill \break
    \indent \phantom{ii)\ }  $m_i(M_{GUT}) \equiv m_0$,

    iii) Universality of the trilinear scalar couplings:
         $A^{l}(M_{GUT}) = A^{d}(M_{GUT}) = A^{u}(M_{GUT})$ \hfill \break
    \indent \phantom{iii)\ }  $\equiv A_0 m_0$. \hfill\break

As extensively discussed in Ref. \cite{bbefms1}, these conditions have strong 
consequences for low--energy supersymmetry
phenomenology, and in particular for the properties of the neutralino, which
is defined as the lowest--mass linear superposition of the
two neutral gauginos ($\tilde \gamma$ and $\tilde Z$) and the two
neutral higgsinos ($\tilde H_1$ and $\tilde H_2$):
\begin{equation}
\chi = a_1 \tilde \gamma + a_2 \tilde Z + a_3 \tilde H_1 + a_4 \tilde H_2\;.
\end{equation}
\noindent
Different neutralino compositions are classified in terms of the 
parameter $P \equiv a_1^2 + a_2^2$: {\it gaugino--like} when $P > 0.9$,
{\it mixed} when $0.1 \leq P \leq 0.9$ and {\it higgsino--like} when $P < 0.1$. 

The unification conditions represent a theoretically attractive
possibility, which makes strictly universal SUGRA models very predictive.
However, the above assumptions, particularly ii) and iii), are not fully justified, 
since universality may occur at a scale higher
than $M_{GUT}$, i.e. the Planck scale or string scale, 
in which case renormalization above $M_{GUT}$ weakens universality in the
$m_i$. Deviations from some of the
unification conditions have been considered by a number of authors 
\cite{bbefms1,others,bbefms2}. Implications of these deviations for relic
neutralino phenomenology have been  discussed in detail in Refs. 
\cite{bbefms1,bbefms2}. 

In the present paper, we discuss the DAMA/NaI data both in a SUGRA model 
with strict unification conditions and in a SUGRA framework, where 
we introduce a departure from universality in the scalar masses at $M_{GUT}$ 
which splits the Higgs mass parameters $M_{H_1}$ and $M_{H_2}$ in the following way:
\begin{equation}
M^2_{H_i}(M_{GUT}) = m_0^2(1+\delta_i)~.
\label{eq:nonuniv}
\end{equation}
The parameters $\delta_i$ which quantify the departure from universality for
the $M^2_{H_i}$ will be varied in the range ($-1$,$+1$), but are taken to be
independent of the other supersymmetric parameters. 

Our supersymmetric parameter space is constrained by the following conditions: 
(a) all experimental bounds on Higgs, neutralino, chargino and sfermion masses
are satisfied (for current LEP bounds see, 
for instance, Ref. \cite{lep2,ichep}), (b) the neutralino is
the Lightest Supersymmetric Particle (LSP), (c) the constraints on the $b
\rightarrow s + \gamma$ process are satisfied, 
(d) the constraints on the mass of the bottom quark $m_b$ are also satisfied, 
however with a 
$b-\tau$ Yukawa unification relaxed by about $20\%$, (e) EWSB is realized
radiatively, (f) the neutralino relic abundance $\Omega_{\chi} h^2$ does not
exceed the cosmological upper bound, which is conservatively set here as 
$\Omega_{\chi} h^2 \leq 0.7$. Because of the requirements of radiative EWSB and
of the universality conditions on the gaugino masses and on the trilinear
couplings, the independent supersymmetric parameters are reduced to the
following set (apart from the $\delta_i$'s): $m_{1/2}, m_0, A_0, \tan \beta=v_2/v_1$
and ${\rm sign}(\mu)$.

The Renormalization Group Equations (RGE's) are solved by 
using the 1--loop  beta functions
including the whole supersymmetric particle spectrum
from the GUT scale down to $M_Z$,
neglecting the possible effects of intermediate thresholds.
Two--loop and threshold effects on the running of the gauge and Yukawa
couplings are known not to exceed 10\% of the final result \cite{lang}.
While this is
of crucial importance as far as gauge coupling unification is
concerned \cite{lang}, it is a second--order
effect on the evolution of the soft masses and then it is neglected here. 

In order to specify the supersymmetry phenomenology,
boundary conditions for the gauge and Yukawa couplings have to be
specified.
Low--scale values for the gauge couplings and for the top--quark and the
tau--lepton Yukawa couplings are fixed using present experimental results.
In particular, we assign for the top mass the value $m_t=175$ GeV. 

A few qualifications are in order here about the constraints due to the 
$b \rightarrow s + \gamma$ process and for the bottom mass. 
In our analysis, the inclusive decay rate 
BR($B \rightarrow X_s \gamma$) 
\cite{bertolini,bg,garisto,borzumati,wu,barger} 
is calculated with corrections up to the 
leading order. Next--to--leading order corrections 
\cite{chetyrkin,ciuchini1,cza,ciuchini2} 
are included only when 
they can be applied in a consistent way, i.e. both to standard model 
and to susy diagrams. This criterion limits the use of 
next--to--leading order corrections to peculiar regions of the supersymmetric 
parameter space, where the assumptions, under which the next--to--leading order 
susy corrections have been obtained, apply \cite{ciuchini2}. 
We require that our theoretical evaluation for BR($B \rightarrow X_s \gamma$) 
is within the  range:  
1.96 $\times 10^{-4} \leq$ BR($B \rightarrow X_s \gamma$) $\leq$ 4.32
$\times 10^{-4}$. This range is obtained by combining the experimental 
data of Refs. \cite{glenn,barate} at 95\% C.L. and by adding a 
theoretical uncertainty of 25\%, whenever the still incomplete 
next--to--leading order susy corrections cannot be applied.

The supersymmetric corrections to the bottom mass include
contributions from bottom--squark--gluino loops and from
top--squark--chargino loops \cite{carena}.
In the present analysis, the bottom mass is computed as a function of the other
parameters and required to be compatible with the present experimental bounds.
Theoretical uncertainties in the evaluation of $m_b$ arise from the
running of the RGE's. 
Since our choice
is to solve RGE's at the 1--loop level and without thresholds,
we estimate an uncertainty of the order of 10\% in our prediction for
$m_b$. To take into account such an 
uncertainty we have chosen to weaken the bounds on $m_b$ given in
\cite{ceop} by 10\%. Thus we require  $m_b$
to fall into the range $2.46~ \mbox{GeV} \leq m_b(M_Z) \leq 3.42~ \mbox{GeV}$,
 at 95\% C.L. As mentioned above, 
$b-\tau$ Yukawa unification is relaxed by about $20\%$.

The neutralino relic abundance $\Omega_{\chi} h^2$ is 
calculated as illustrated in 
\cite{ouromega}. As already stated, we apply 
to our supersymmetric 
parameter space an upper bound conservatively set at the 
value $\Omega_\chi h^2 \leq 0.7$, and we consider as cosmologically 
interesting the range $0.01 \leq \Omega_{\chi} h^2 \leq 0.7$. However, 
we stress that, according to the most recent data and analyses 
\cite{omegamatter}, the most appealing interval for the 
neutralino relic abundance is the narrower one: 
$0.02 \lsim \Omega_{\chi} h^2 \lsim 0.2$. The local neutralino density
$\rho_\chi$ is factorized in terms of the total local 
dark matter density $\rho_l$ as $\rho_\chi = \xi \rho_l$.
The parameter $\xi$ is calculated according to the usual rescaling
recipe \cite{gaisser}:
$\xi = {\rm min}(\Omega_{\chi} h^2/(\Omega h^2)_{\rm min},1)$. We take here
$(\Omega h^2)_{\rm min} = 0.01$.

In our analysis the $m_{1/2}, m_0, A_0, \tan \beta$  parameters 
are varied in
the following ranges: $10\;\mbox{GeV} \leq m_{1/2} \leq  500\;\mbox{GeV},\;
 m_0 \leq  1\;\mbox{TeV},\;
-3 \leq A_0 \leq +3,\;
1 \leq \tan \beta \leq 50$; the parameter $\mu$ is taken positive.
We remark that the values taken here as upper limits of the ranges for 
$m_{1/2}, m_0$ are inspired by the upper 
bounds which may be
derived for these quantities in SUGRA theories, when one requires that the 
EWSB, radiatively induced by the soft supersymmetry
breaking, does not occur with excessive fine tuning 
(see Ref. \cite{bbefms1} and references quoted therein).

\section{Rate for WIMP direct detection}

The indication of a possible annual modulation effect, singled out by 
the DAMA/NaI data \cite{dama2}, points, at a 2--$\sigma$ C.L., to a 
very delimited region in  the plane 
$\xi \sigma^{(\rm nucleon)}_{\rm scalar}$ -- $m_\chi$, 
where $m_\chi$ is the WIMP mass and
$\sigma^{(\rm nucleon)}_{\rm scalar}$ is the WIMP--nucleon elastic scalar 
cross section.

When the uncertainties in $\rho_l$ are taken into account, 
the original 2--$\sigma$ C.L. region singled out by DAMA/NaI data has to be 
enlarged into a region $R$, where the quantity 
$\xi \sigma^{(\rm nucleon)}_{\rm scalar}$ falls into the 
following range \cite{noi}: 
$(1 - 3) \times 10^{-9}\; {\rm nb} \lsim 
\xi \sigma^{(\rm nucleon)}_{\rm scalar} \lsim 3 \times 10^{-9}\; {\rm nb}$, 
in the mass range 
30 GeV $\lsim m_{\chi} \lsim$ 110 GeV. 

In Ref. \cite{noi}, 
where the relevant formulae for the 
evaluation of $\sigma^{(\rm nucleon)}_{\rm scalar}$ are reported, it is also shown that,  
in the MSSM,  wide domains of the supersymmetric parameter space 
provide values for $\sigma^{(\rm nucleon)}_{\rm scalar}$ and $m_{\chi}$  
which are  within region $R$. 
Out of the two competing contributions to $\sigma^{(\rm nucleon)}_{\rm
scalar}$, Higgs--exchange 
 and squark--exchange processes, usually the former dominates over 
the latter one. 
Then let us discuss which properties 
of the Higgs--exchange amplitude are instrumental in making this 
contribution sizeable enough as required by the modulation 
effect: 
$\xi \sigma_{\rm scalar}^{(\rm nucleon)} \gsim 1 \times 10^{-9}\; {\rm nb}$. 
 As is clear from the expressions given in 
Ref. \cite{noi}, the most important  parameters for establishing the 
size of the Higgs--exchange amplitude are:  $m_h$, 
 $\tan \beta$, and the mixing angle $\alpha$ of the two 
CP--even neutral Higgs bosons ($h$ and $H$). The largest values for 
the Higgs--exchange amplitude occur for the following combination 
of their respective values: small $m_h$ (because of the propagator), 
large values of $\tan \beta$ and of $\alpha$ 
(because of the structure of the Higgs--quark couplings). 
In SUGRA these three parameters are rather strongly correlated \cite{bbefms1},
 so that requiring some lower bound $\xi \sigma_{\rm scalar}^{(\rm nucleon)}$ 
imposes severe constraints on the parameter space. These 
properties may be suitably discussed in terms of the mass $m_A$ of the CP--odd
neutral Higgs boson (however, we remind that this parameter is not free, but
depends on the parameters defining our parameter 
space and on the RGE's evolution). 

{From} Fig. 1, which displays a generic scatter plot of 
$\xi \sigma_{\rm scalar}^{(\rm nucleon)}$ 
versus $m_A$ in the universal SUGRA model (i.e. with $\delta_i$'s = 0), 
we see that the 
lower bound 
$\xi \sigma_{\rm scalar}^{(\rm nucleon)} \gsim 1 \times 10^{-9}\; {\rm nb}$ 
implies 
the upper bound $m_A \lsim 180$ GeV. 
If we further take into account the scatter plot of $m_A$ in terms of 
$\tan \beta$,  shown in Fig. 2, we obtain in turn a lower limit on $\tan
\beta$: $\tan \beta \gsim 42$. We also note that just a small relaxation in the
lower bound for 
$\xi \sigma_{\rm scalar}^{(\rm nucleon)} \gsim 1 \times 10^{-9}\; {\rm nb}$  
would also allow intermediate values of $\tan \beta$: 
$\tan \beta \sim 5~-~10$. The reasons for the typical feature of the scatter
plot of Fig. 2 for $m_A \lsim 300$ GeV are discussed in the Appendix. 

If we now require  our supersymmetric configurations to lie inside the
modulation region $R$ (this set of configurations is denoted as set $S$),
we obtain the scatter plot of Fig. 3, which proves that 
the annual modulation data are in fact compatible with a universal SUGRA
scheme.  Fig. 4 shows that a number of the selected supersymmetric 
configurations 
 fall  into the cosmological interesting range of 
$\Omega_{\chi} h^2$. 

The other qualifications for the configurations which lie inside the 
region $R$, relevant for
searches at accelerators, concern the ranges for the $h$--Higgs boson mass, 
the neutralino mass and the lightest top--squark mass, which we find to be:
$m_h \lsim 115$ GeV, 50 GeV $\lsim m_{\chi} \lsim$ 100 GeV and 
200 GeV $\lsim m_{\tilde t_1} \lsim$ 700 GeV, respectively. 

Let us now turn to SUGRA schemes with deviations from universal scalar masses
(i.e. $\delta_i$'s $\not= 0$). By varying the usual supersymmetric parameters 
as before and the $\delta_i$'s in the range $-1 \leq \delta_i \leq +1$ we find 
the scatter plot of Fig. 5, which shows that 
the requirement 
$\xi \sigma_{\rm scalar}^{(\rm nucleon)} \gsim 1 \times 10^{-9}\; {\rm nb}$ 
implies now 
a more relaxed upper bound on $m_A$: $m_A \lsim 330$ GeV. {From} Fig. 6, 
we see that this upper limit on $m_A$ is compatible with all values of 
$\tan \beta$. 
In Figs. 7 and 8 we notice that, as expected, our new sample of representative 
points covers a slightly wider domain of region $R$, and, more significantly, 
contains new  neutralino configurations of cosmological interest. Figs. 9--11 
show how the values for the 
$h$--Higgs boson mass, the neutralino mass and the 
lightest top--squark mass are
distributed in terms of $\tan \beta$. The ranges of these masses
are similar to those already found in the universal case,
but now $\tan\beta$ extends to the interval $10 \lsim \tan \beta \lsim 50$,
instead of being limited only to very large values.

The distribution of the values for the parameters $\delta_i$'s, which provide 
supersymmetric configurations in agreement with the annual modulation data, 
are shown in Fig. 12. The peculiar distribution of representative points 
in the left--upper side of the figure may be easily 
understood in terms of the general 
properties discussed in Ref. \cite{bbefms1}.
The generic trend displayed in this figure shows that small values
of $\tan\beta$ require sizeable deviations from universality in the Higgs
mass parameters.

\section{Conclusions}

We have analysed the total sample of new and former DAMA/NaI data 
\cite{dama2,dama1}, which provide the indication of a possible annual
modulation effect in the rate for WIMP direct detection. We have demonstrated 
that these experimental data, already proved to be widely compatible 
with relic neutralinos of cosmological interest in a MSSM
scheme \cite{noi,noi0}, are also compatible with a SUGRA framework.

We have specifically considered a supergravity  scheme with strict unification 
conditions on scalar and gaugino masses and on trilinear scalar couplings,
and supergravity models with departures from universality in the scalar mass
parameters of the Higgs sector. We have proved that in
universal SUGRA neutralino configurations with interesting cosmological
properties may be found. Even more so, in case of non--universal SUGRA models. 

Other main results of our analysis are the following.  
In the universal SUGRA model the constraints imposed by the DAMA/NaI data 
imply for the $h$--Higgs boson mass, 
the neutralino mass and the lightest top--squark mass, the following ranges: 
$m_h \lsim 115$ GeV, 50 GeV $\lsim m_{\chi} \lsim$ 100 GeV and 
200 GeV $\lsim m_{\tilde t_1} \lsim$ 700 GeV, respectively. 
In universal SUGRA $\tan \beta$ is constrained to be large, 
 $\tan \beta \gsim 42$, whereas, 
 with departure from universality in the scalar masses, the range for 
$\tan \beta$ widens to $10 \lsim \tan \beta \lsim 50$. 

\section{appendix}
\appendix

In this Appendix we show  that the peculiar behaviour of the scatter plot
in Fig. 2 for $m_A \lsim 300$ GeV is induced, in the universal SUGRA model, 
by the combined effect of the limits on the  $b\rightarrow s+\gamma$ decay 
rate and on the mass of the bottom quark $m_b$. Let us start by reminding  
that in generic SUGRA models $m_A$ is a function
of the other parameters through the radiatively induced EWSB mechanism. 
A useful parametrization is \cite{bbefms1}:
\begin{equation}
m_A^2= K_1 m_{1/2}^2+ K_2 m_0^2 + K_3 A_0^2 m_0^2 + K_4 A_0
m_0 m_{1/2}-m_Z^2,
\label{eq:ma_rge}
\end{equation}
where the coefficients $K_i$ are only functions of $\tan\beta$
and the $\delta_i$'s. Their properties are extensively discussed
in Ref. \cite{bbefms1}, to which we refer for details. 
To simplify the discussion let us take $A_0 = 0$. 

In the universal SUGRA model 
the coefficients $K_1$ and $K_2$ are both decreasing functions of $\tan\beta$, 
being $K_1$ $\sim$ 3 and $K_2 \sim$ 1 for $\tan\beta \sim$ 5--10
and both vanishingly small at large 
$\tan\beta$. Actually, while $K_1$ is always positive, $K_2$ becomes
negative for $\tan\beta \sim$ 50. 
Furthermore $m_{1/2}$ is limited from above by the annual modulation data which imply
$m_{\chi}$$\lsim$ 110 GeV and then
$m_{1/2}\sim 2.5\; m_{\chi} \lsim 270$ GeV (the relation 
$m_{1/2}\sim 2.5\; m_{\chi}$ holds, since neutralino is mainly gaugino--like in
universal SUGRA). 
Instead, for $m_0$ we have a lower bound which depends on $\tan\beta$,
as displayed in Fig. A.1. This bound arises as a combined effect of the 
$b\rightarrow s+\gamma$ and $m_b$ constraints.
Consequently $m_A$ may be small only for $\tan\beta \lsim 10$, where $m_0$ may be arbitrarily 
small, or at large $\tan\beta$, where the product $K_2 m^2_0$ is kept small by $K_2$.  

In SUGRA models with $\delta_i \neq 0$ the coefficient $K_2$
may become vanishingly small also for intermediate values of $\tan\beta$
so that, in non--universal models,  small values of $m_A$ may occur over the whole
$\tan\beta$ range, as shown in Fig. 6.

\vfill
\eject

\begin{center}
{\large FIGURE CAPTIONS}
\end{center}
\vspace{1cm}

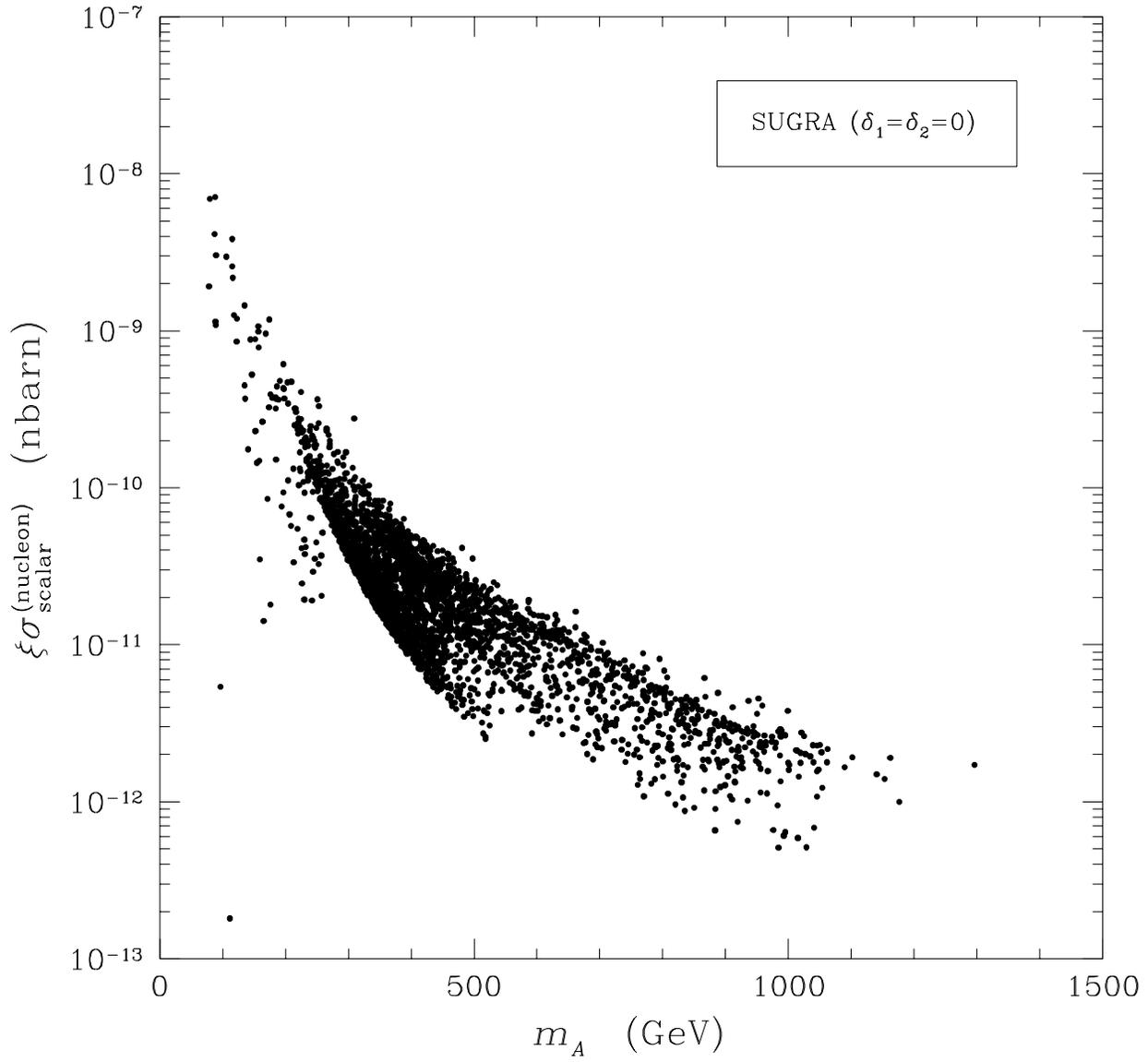
{\bf Figure 1} --
Scatter plot of $\xi \sigma_{\rm scalar}^{(\rm nucleon)}$ versus $m_A$ 
for a scanning of the supersymmetric 
parameter space as defined in Sect. II, in universal SUGRA. 

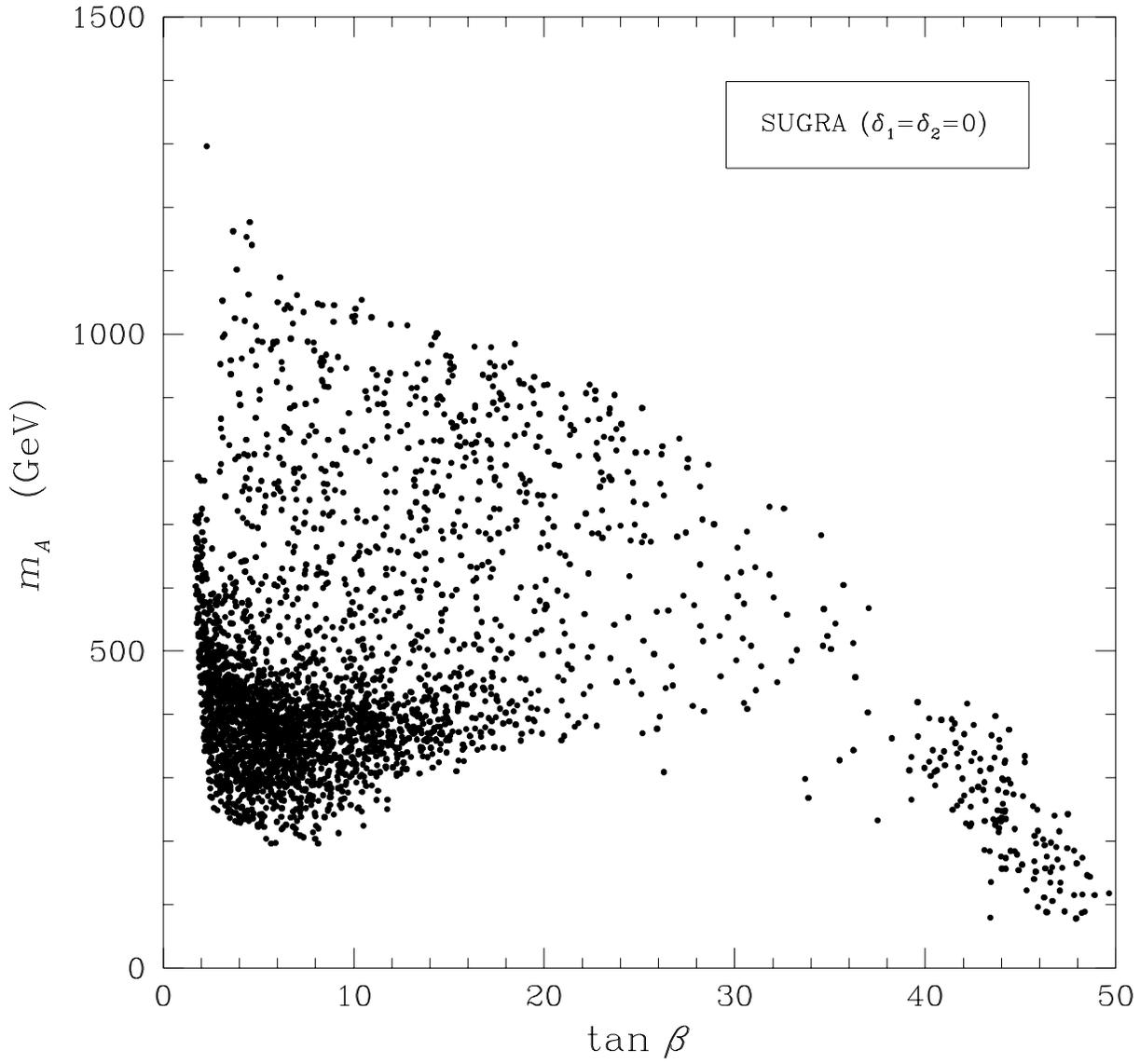
{\bf Figure 2} --
Scatter plot of $m_A$ versus $\tan \beta$ 
for a scanning of the supersymmetric 
parameter space as defined in Sect. II, in universal SUGRA. 

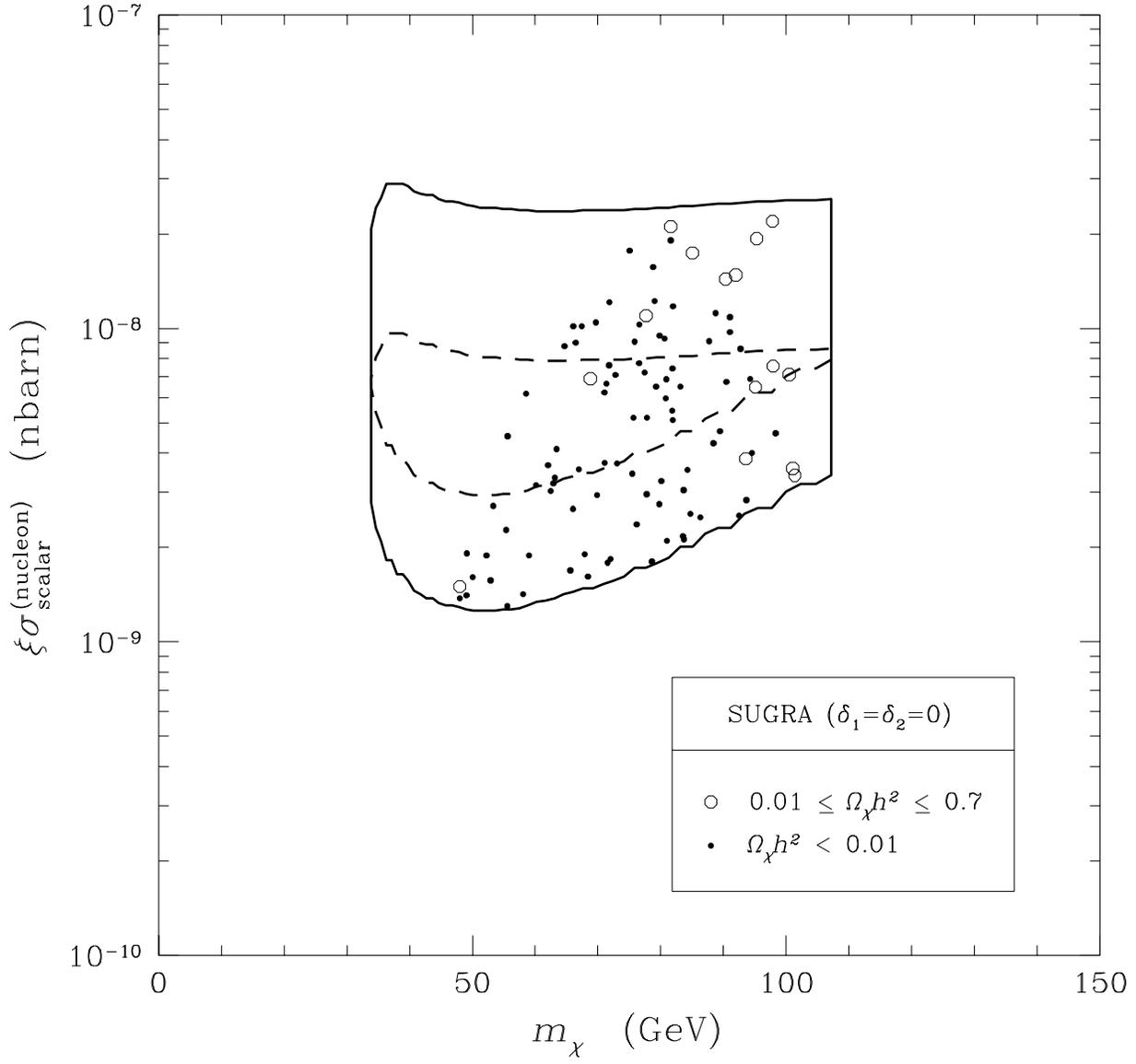
{\bf Figure 3} -- 
Scatter plot of set $S$ in the plane 
$m_{\chi}$--$\xi \sigma_{\rm scalar}^{(\rm nucleon)}$ in universal SUGRA. 
The dashed contour line
delimits the 2--$\sigma$ C.L. region, obtained by the DAMA/NaI Collaboration, 
by combining together the data of
the two running periods of the annual modulation experiment \cite{dama2}.  
The solid contour  line is obtained from the dashed line, which refers to the
value $\rho_l = 0.3 ~$ GeV cm$^{-3}$, by accounting for 
the uncertainty range of $\rho_l$, as explained in Sect. III (the region 
delimited by the solid line is denoted as region $R$ in the text). 
The representative points are denoted differently depending on 
the values of the neutralino relic abundance $\Omega_{\chi} h^2$.

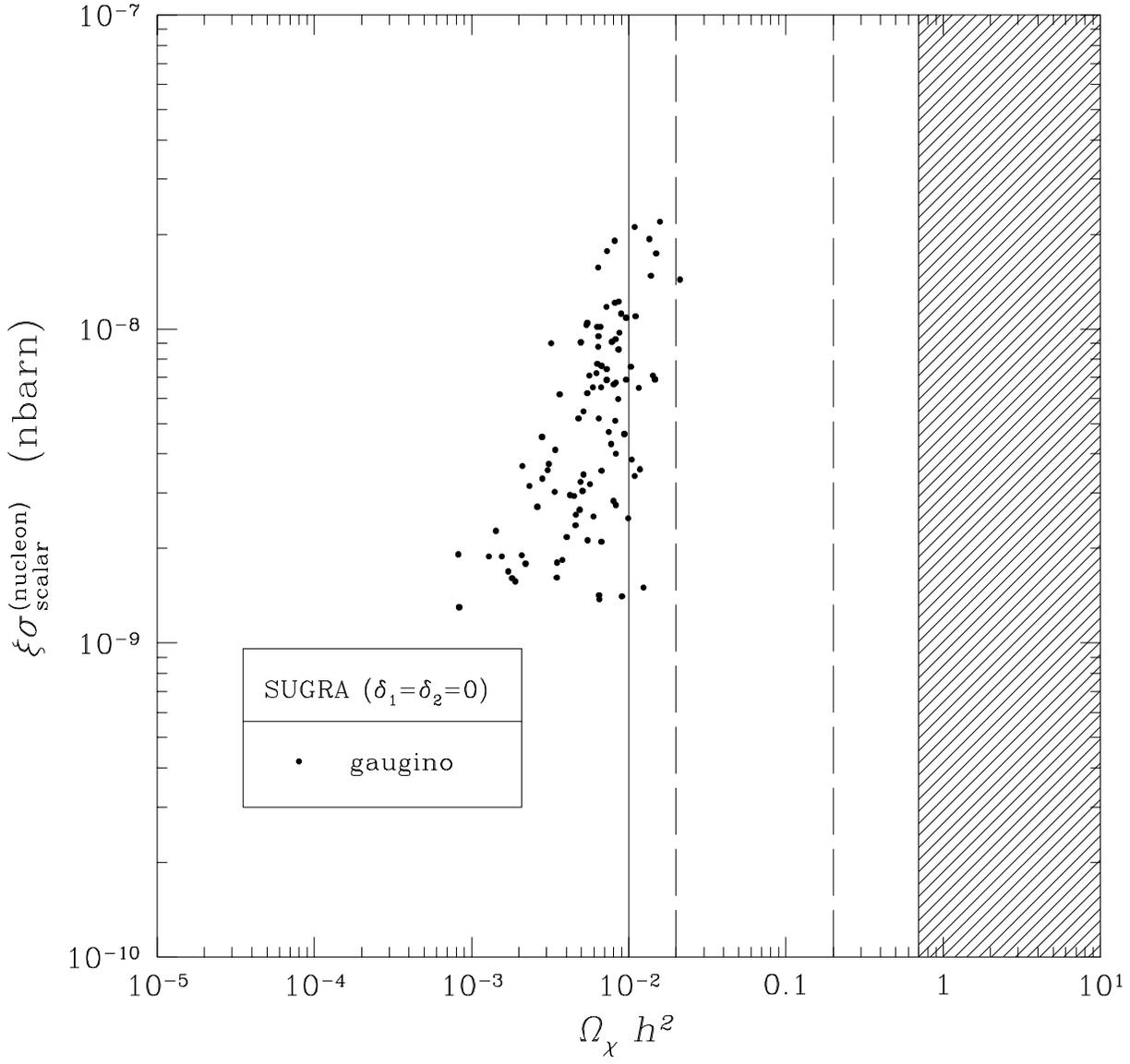
{\bf Figure 4} -- Scatter plot of set $S$ in the plane 
 $\Omega_{\chi} h^2$ -- $\xi \sigma_{\rm scalar}^{(\rm nucleon)}$ in 
universal SUGRA. Here neutralinos turn out to be gaugino--like only. 
The two vertical solid lines 
delimit the $\Omega_{\chi} h^2$ range of cosmological interest. The two
dashed lines delimit the most appealing interval for 
$\Omega_{\chi} h^2$, as suggested by the most recent observational data.

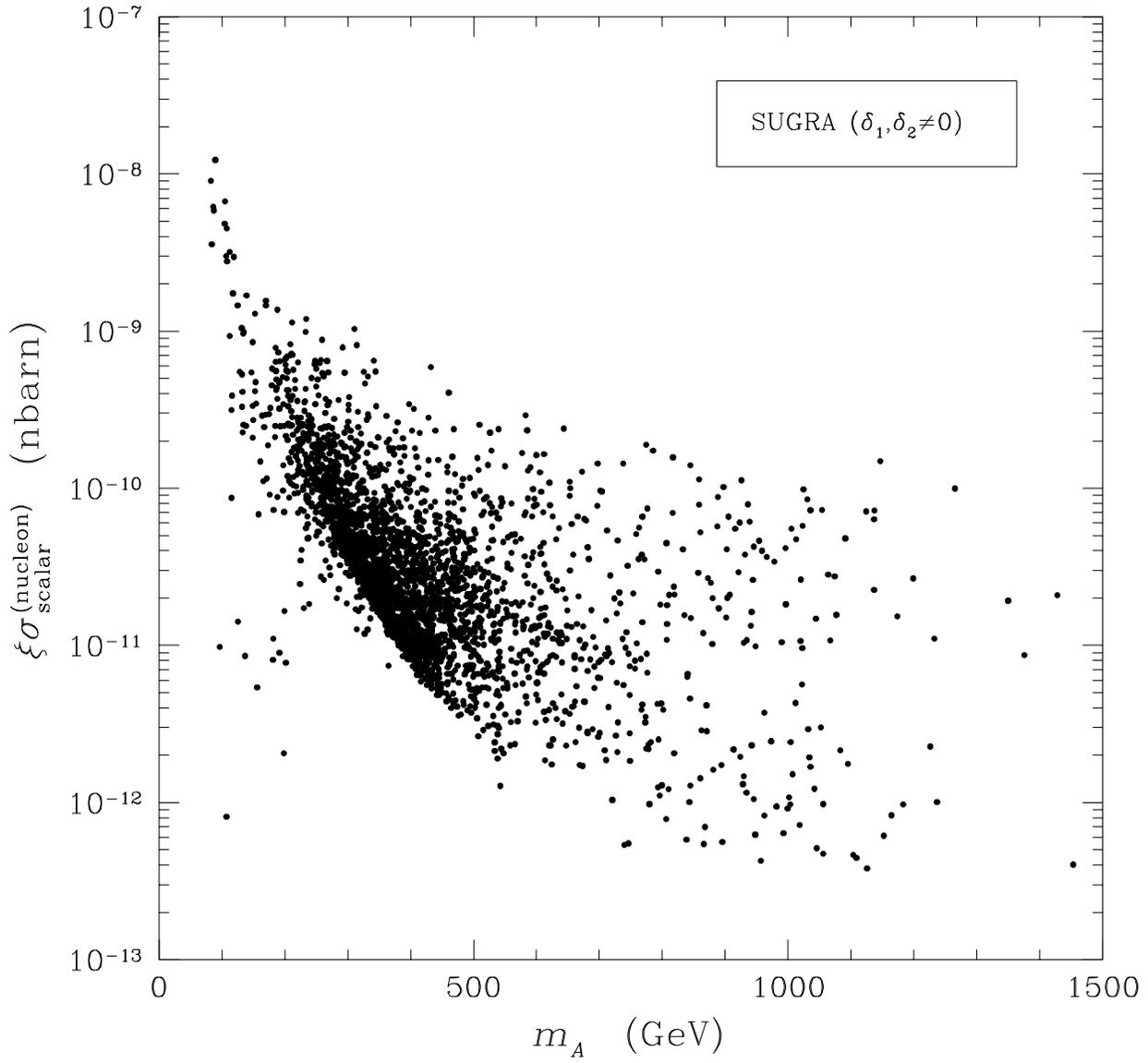
{\bf Figure 5} --
Scatter plot of $\xi \sigma_{\rm scalar}^{(\rm nucleon)}$ versus $m_A$ 
for a scanning of the supersymmetric 
parameter space as defined in Sect. II, in non--universal SUGRA.

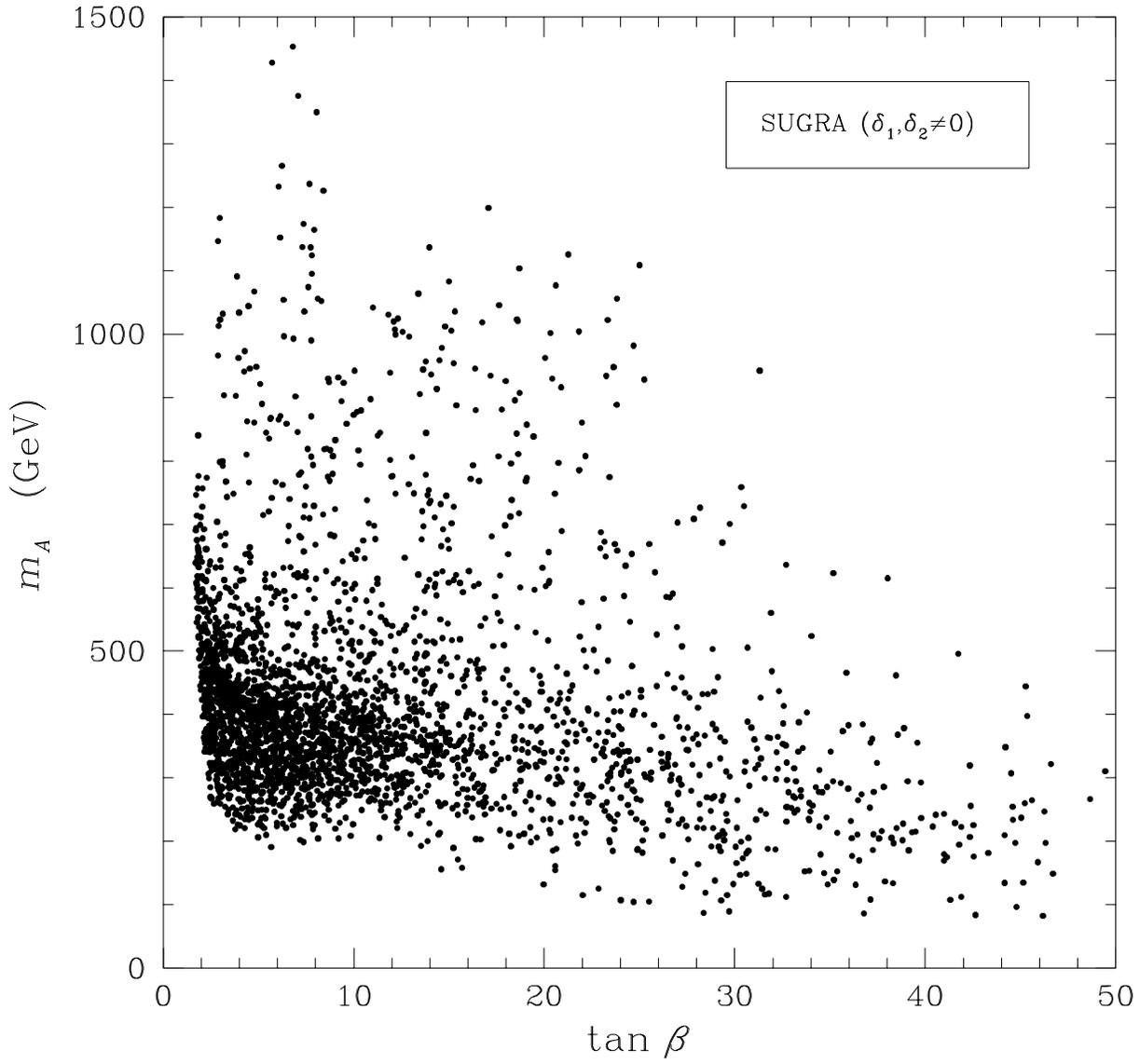
{\bf Figure 6} --
Scatter plot of $m_A$ versus $\tan \beta$ 
for a scanning of the supersymmetric 
parameter space as defined in Sect. II, in non--universal SUGRA. 

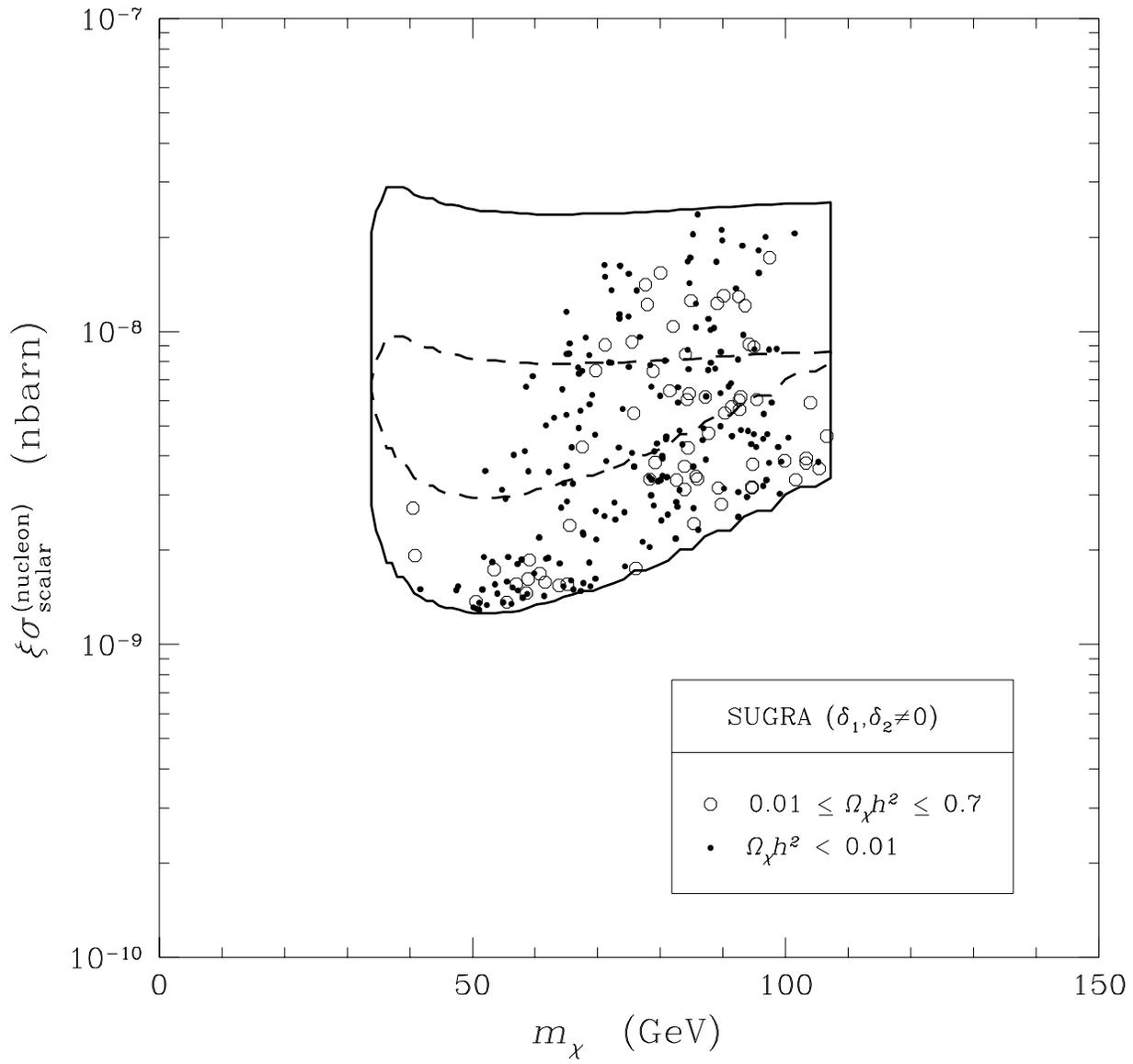
{\bf Figure 7} --
As in Fig. 3, for non--universal SUGRA. 

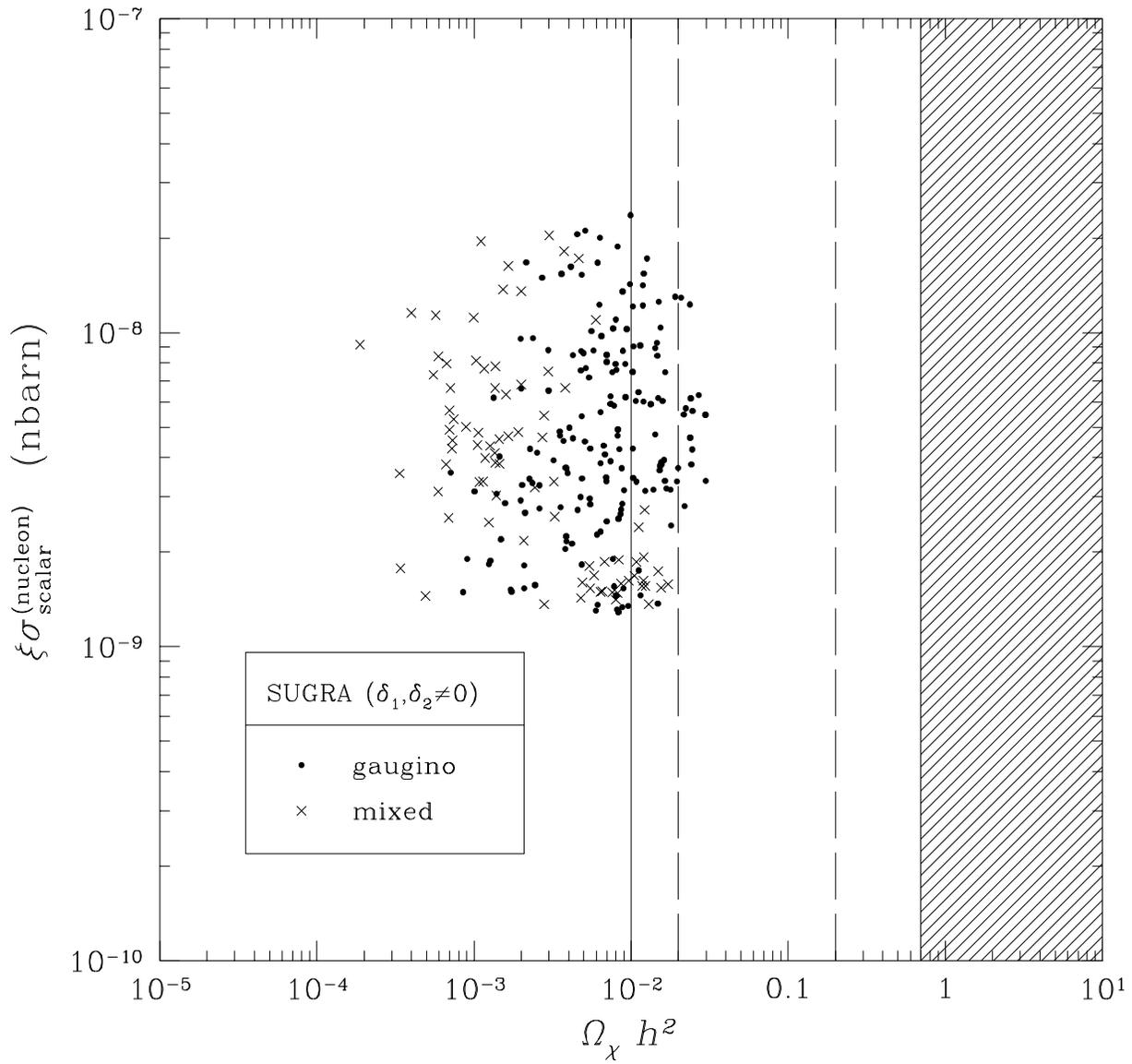
{\bf Figure 8} --
As in Fig. 4, for non--universal SUGRA. Here dots and crosses 
denote neutralinos of different composition according to the
classification in Sect. II. 

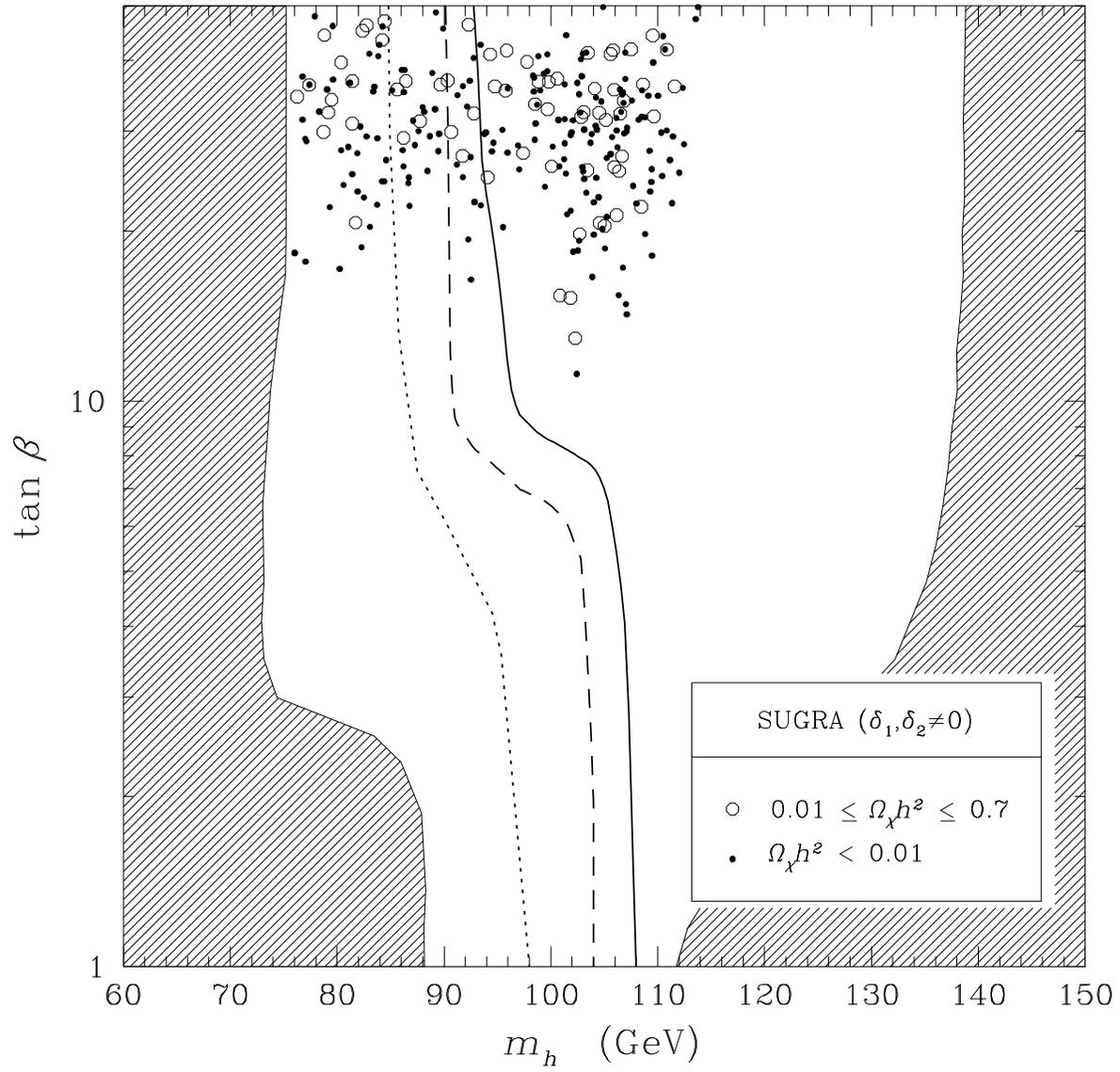
{\bf Figure 9} --
Scatter plot of set $S$ in the plane $m_h$ -- $\tan \beta$, in 
non--universal SUGRA. The hatched region on the right is excluded by theory. 
The hatched region on the left is 
excluded by present LEP data at $\sqrt s$ = 183 GeV. The dotted and the dashed 
curves denote the reach of LEP2 at energies $\sqrt s$ = 192 GeV and 
$\sqrt s$ = 200 GeV, respectively. The solid line represents the 
95\% C.L. bound reachable at LEP2, in case of non discovery of a neutral 
Higgs boson. 

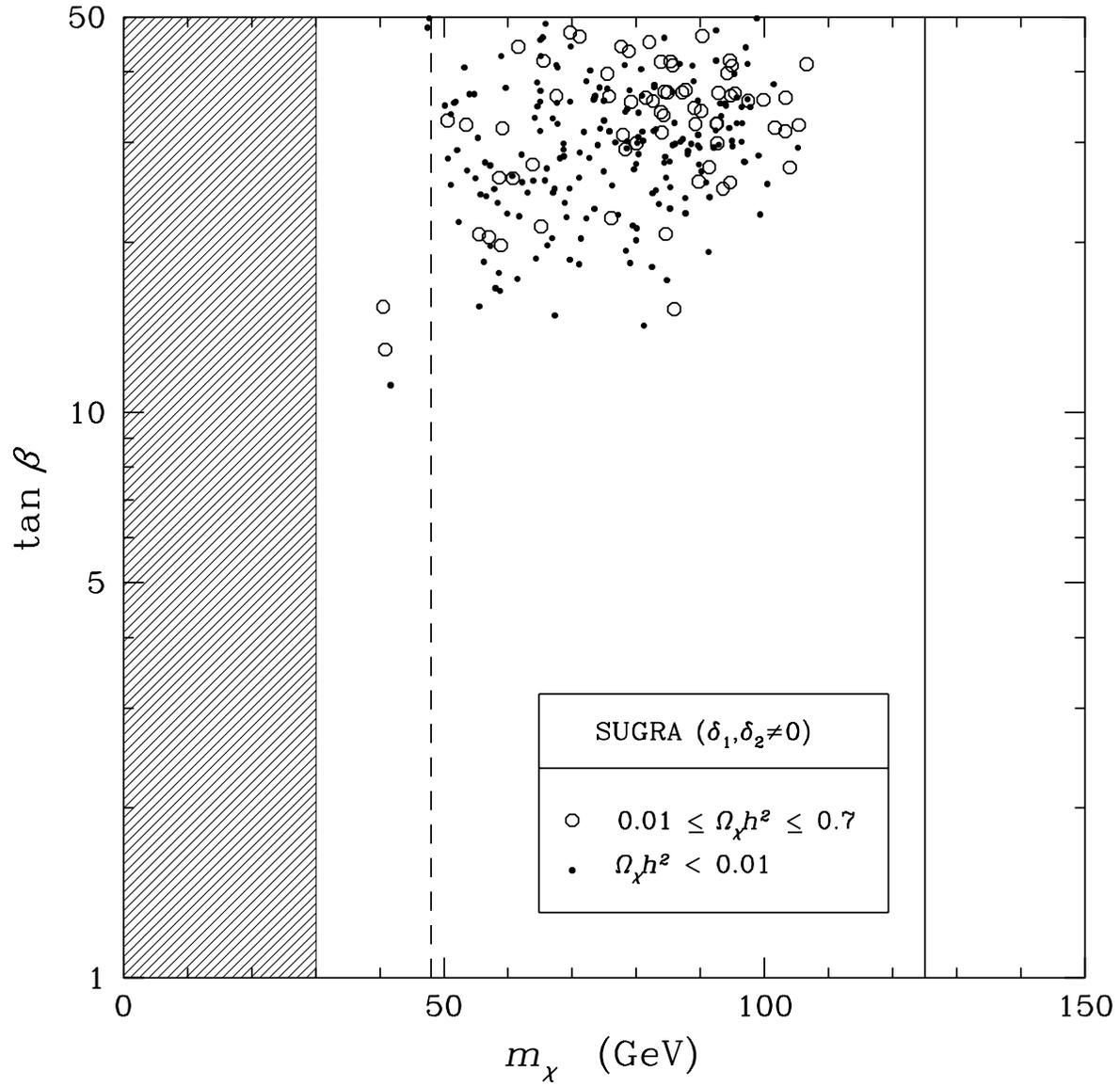
{\bf Figure 10} -- Scatter plot of set $S$ in the plane 
$m_{\chi}$ -- $\tan \beta$, in non--universal SUGRA.
The hatched region on the left is 
excluded by present LEP data. The dashed and the 
solid vertical lines denote the reach of LEP2 and TeV33, respectively. 

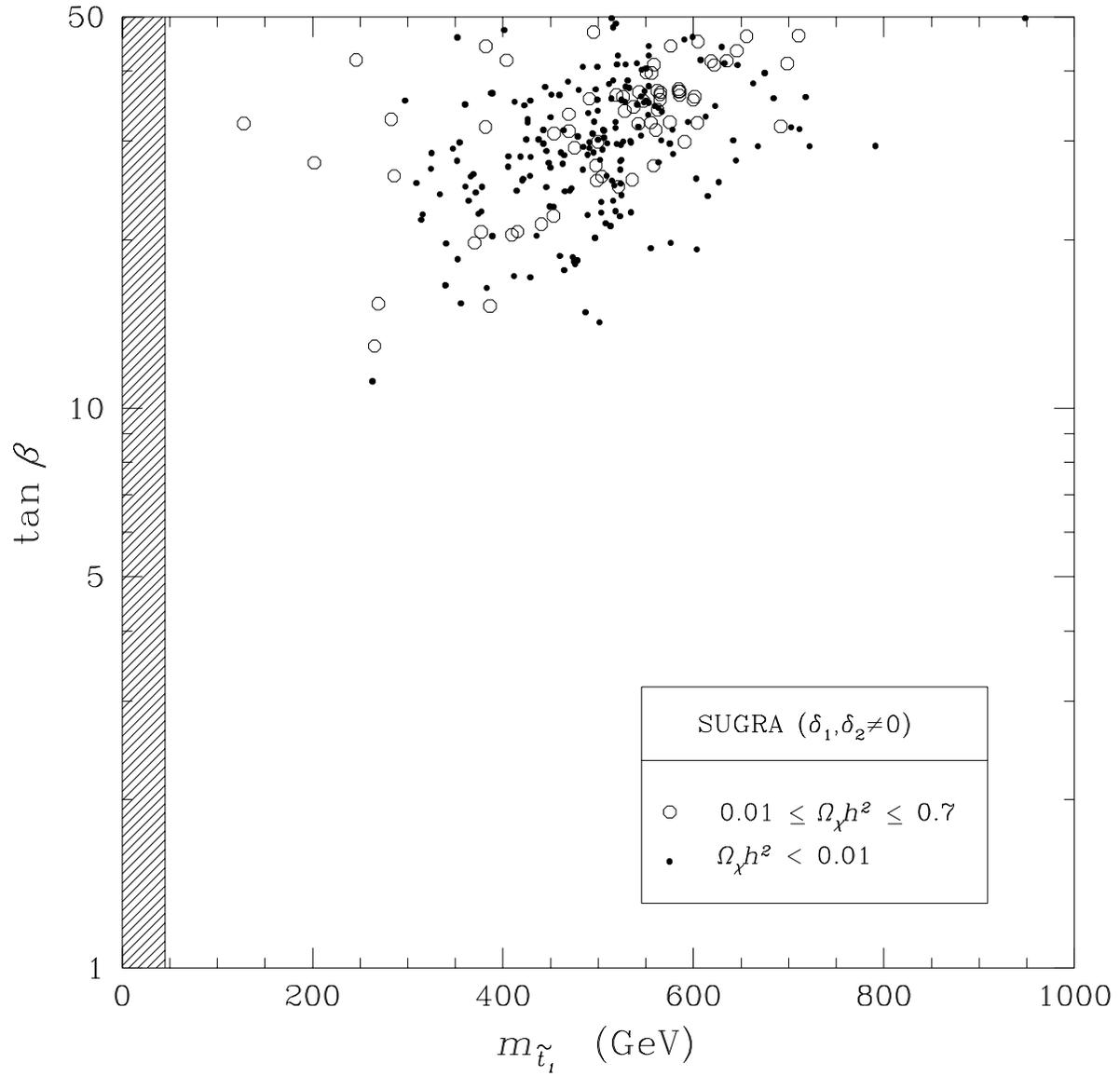
{\bf Figure 11} -- Scatter plot of set $S$ in the plane 
$m_{\tilde t_1}$ -- $\tan \beta$, in non--universal SUGRA.
The hatched region is excluded by LEP data (without any restriction on other 
masses). 

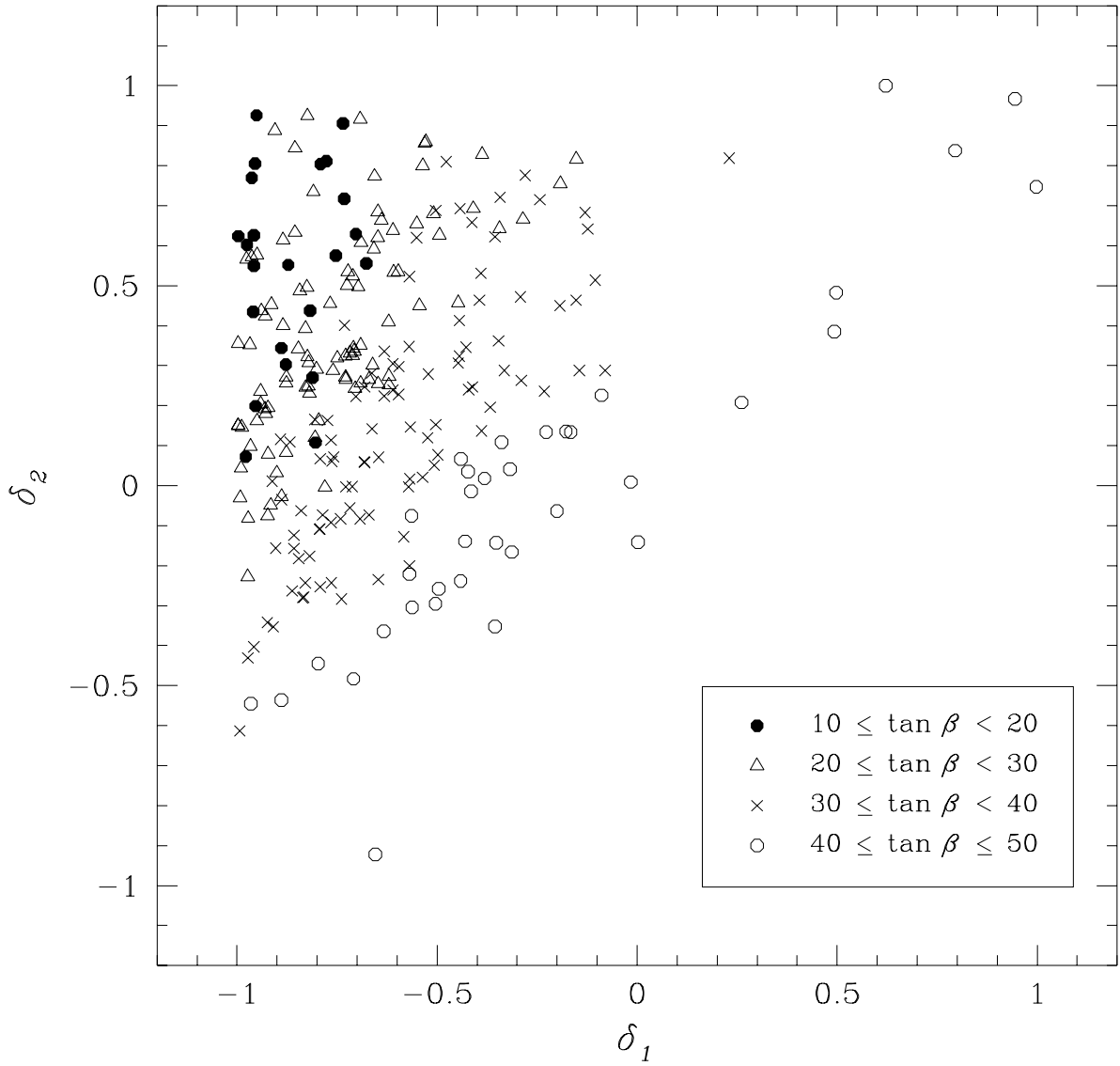
{\bf Figure 12} -- Scatter plot of the values of the $\delta_1$ and $\delta_2$
parameters for set $S$. 

{\bf Figure A.1} -- Scatter plot in the plane $m_0$--$\tan \beta$,
for a scanning of the supersymmetric 
parameter space as defined in Sect. II, in universal SUGRA.

\vfill\eject

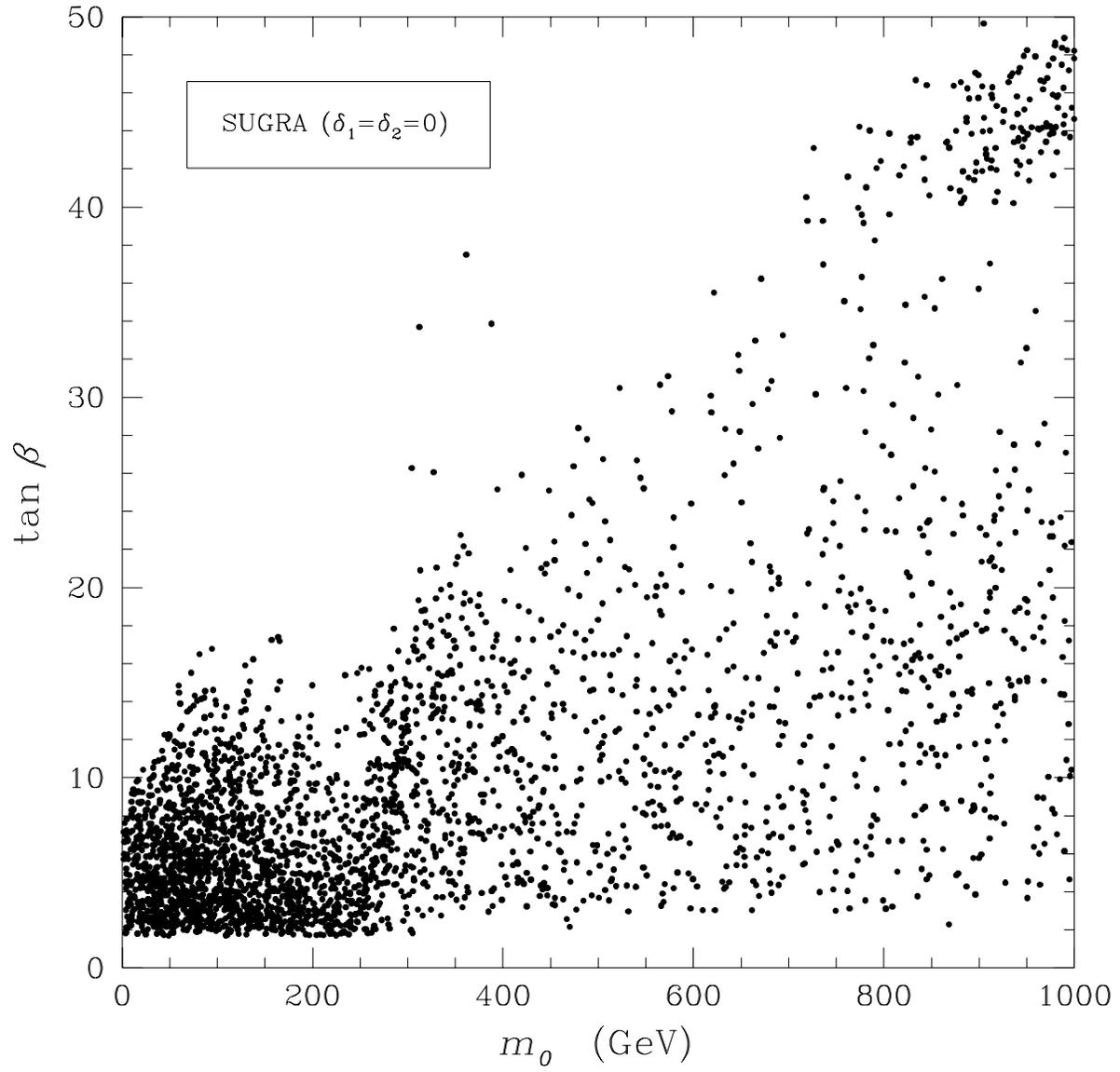
\end{document}